\documentclass[aps,prb,twocolumn,showpacs,superscriptaddress,amsmath,amssymb]{revtex4}
\usepackage{epsfig}
\usepackage{bm}
\usepackage{color}
\begin{document}

\title{Experimental test of reciprocity relations in quantum thermoelectric transport}

\author{J.~Matthews}
\affiliation{Physics Department and Materials Science Institute, University of Oregon, Eugene, Oregon 97403-1274}
\author{F.~Battista}
\affiliation{Physics Department, Lund University, Box 118, S-221 00, Lund, Sweden}
\affiliation{Nanometer Structure Consortium (nmC@LU), Lund University, Box 118, S-221 00, Lund, Sweden}
\author{D.~S\'{a}nchez}
\affiliation{Instituto de F\'{\i}sica Interdisciplinar y Sistemas Complejos IFISC (CSIC-UIB), E-07122 Palma de Mallorca, Spain}
\author{P.~Samuelsson}
\affiliation{Physics Department, Lund University, Box 118, S-221 00, Lund, Sweden}
\affiliation{Nanometer Structure Consortium (nmC@LU), Lund University, Box 118, S-221 00, Lund, Sweden}
\author{H.~Linke}
\affiliation{Physics Department, Lund University, Box 118, S-221 00, Lund, Sweden}
\affiliation{Nanometer Structure Consortium (nmC@LU), Lund University, Box 118, S-221 00, Lund, Sweden}

\begin{abstract} 
  Fundamental symmetries in thermoelectric quantum transport, beyond
  Onsagers relations, were predicted two decades ago but have to date
  not been observed in experiments. Recent works have predicted the
  symmetries to be sensitive to energy-dependent, inelastic
  scattering, raising the question whether they exist in
  practice. Here we answer this question affirmatively by
  experimentally verifying the thermoelectric reciprocity relations in
  a four-terminal mesoscopic device where each terminal can be
  electrically and thermally biased, individually. The linear response
  thermoelectric coefficients are found to be symmetric under
  simultaneous reversal of magnetic field and exchange of injection
  and emission contacts. We also demonstrate a controllable breakdown
  of the reciprocity relations by increasing thermal bias, putting in
  prospect enhanced thermoelectric performance.
 \end{abstract}
 
\maketitle

\section{Introduction}
Symmetry relations are manifestations of fundamental principles and
constitute cornerstones of modern physics. A prominent example is the
Onsager relations \cite{Onsager} between coefficients connecting
thermodynamic fluxes and forces, which derive from the principle of
microreversibility. In the quantum transport regime, Onsagers
relations for electrical resistance \cite{Buttiker1986} have been
observed in multiterminal mesoscopic systems \cite{Benoit1986,
  vanHouten1989}. In addition to the Onsager relations, reciprocity
relations for thermoelectric (TE) transport coefficients were
predicted \cite{Butcher1990,Jacquod2012}: reversing the magnetic field
and simultaneously exchanging the injection and emission contacts is
expected to leave the coefficients invariant.

In addition to their fundamental interest, the reciprocity relations
are of practical importance. On the one hand, the existence of
symmetry relations could simplify the theory of improved, future TE
materials, such as nanoscale, anisotropic \cite{Matthews2012,Zhou} or
hybrid materials \cite{Snyder} where nonlocal effects may play a
role. On the other hand, the absence of symmetries could be equally
important: asymmetric thermopower was recently shown to allow for
improved TE performance \cite{Casati2011,Brandner,Bala} in the maximum
power regime.

However, to date the reciprocity relations have not been tested
experimentally, and the extent to which they can be observed is
unclear. Recent works \cite{Sanchez2011,Saito2011,Entin2012}
theoretically investigated the robustness of magnetic field symmetries
in the thermopower, which are directly related to the thermoelectric
reciprocity relations.  In contrast to Onsagers relations
\cite{Buttiker1988}, it was predicted that inelastic electron
scattering (always present at finite temperature), in combination with
a breakdown of the Wiedemann-Franz law can break the thermopower
symmetries. The Wiedemann-Franz law is known to break down in
low-dimensional structures due to their strongly energy-dependent
density of states \cite{Zianni} - the same property that makes them
interesting candidates for TE-materials \cite{Dressel}.

A fundamental question is thus: can TE reciprocity relations be
observed in practice and can they be controlled in experiment? Such a
test of the TE reciprocity relations requires a multi-terminal normal
conductor where each terminal can be electrically and thermally
biased, individually, while subjected to an applied magnetic field.

\begin{figure}[h]
\centerline{\psfig{figure=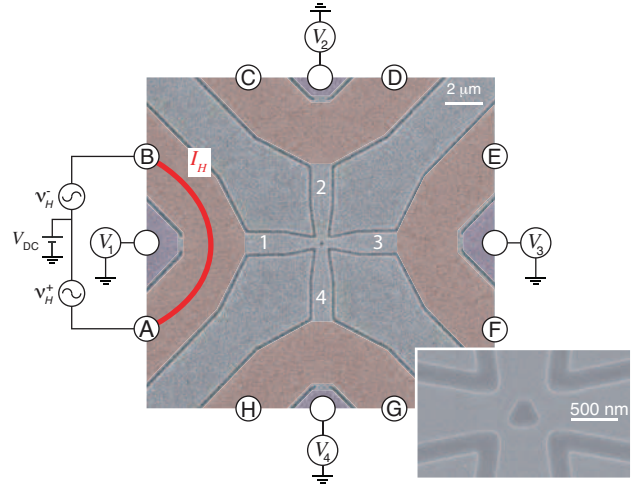}}
\caption{(Color online) Scanning electron micrograph of a device
  identical to the one measured on here, featuring a junction of four
  ballistic micro-channels (terminals) in a cross configuration, with
  an asymmetric scatterer in the central junction. The eight
  surrounding contacts, $\{$A,B,...H$\}$, are used to apply a thermal
  or electrical bias. Four probes are used to measure the terminal
  voltages: $\{V_1, V_2, V_3, V_4\}$. The regions between contact
  pairs, tinted red, can be electrically heated to thermally bias the
  junction. In the configuration shown, the channel between contacts A
  and B are heated through two out-of-phase heating voltages,
  $\nu_H^\pm$ (see Appendix A). (Inset) Close up image of the central region.}
\label{fig:SEMSchematic}
\end{figure}

Here we present such an experimental test in a four-terminal
mesoscopic device (Fig.~\ref{fig:SEMSchematic}), and establish that
the TE reciprocity relations manifest themselves in real devices. We
also find evidence for a breakdown of the relations when we increase
the thermal bias, indicating that the symmetries can be experimentally
controlled, either by inelastic scattering or by non-linear thermal
transport, analogous to the symmetry-breakdown in purely electronic
transport at finite voltages in mesoscopic systems \cite{Linke1998,
  Sanchez2004, Spivak2004,Marlow2006, Leturcq2006, Zumbuhl2006,
  Angers2007, Hartmann2008}. This motivates further investigations on
the symmetry breaking properties and relative role of inelastic
scattering and non-linear thermal transport.

\section{System and method}
We first spell out the properties of the four terminal device. It was
defined by patterning the two-dimensional electron gas (2DEG) formed
in an InP/Ga$_{0.23}$In$_{0.77}$As heterostructure by using
electron-beam lithography and shallow wet etching (for details, see
Ref.~\cite{Matthews2012}). The wafer has a carrier concentration of
$n=1.1\times10^{12}$ cm$^{-2}$ and an electron mobility of
$\mu=3.2\times10^5$ cm$^2$/Vs at 4.2 K. Moreover, all measurements
were made in a He3 cryostat at a background temperature of $\theta_0 =
240$ mK. To create a thermal bias, we used heating voltages of
typically $V_H = 400\mu$V (unless stated otherwise), resulting in
heating currents of less than $400$ nA and an estimated electric
heating power of less than $0.1$ nW delivered to the heating area,
resulting in an estimated temperature rise $\theta < 1$ K. The stray
heating power due to thermal conductance to neighboring heating pads
was negligible. Moreover, we checked carefully that the electric bias
measurements were in the linear response regime (see Appendix A for
details on device properties).

We proceed by defining the thermoelectric coefficients and their
expected symmetry, and describe how they can be determined in
experiments. The linear response of the electrical current flowing in
the $\alpha$'th terminal, $I_\alpha$, of a multi-terminal, mesoscopic
junction is
\begin{equation}
I_{\alpha}=\sum_{\beta\neq \alpha}\left[G_{\alpha\beta}(V_{\alpha}-V_{\beta})+L_{\alpha\beta}(\theta_{\alpha}-\theta_{\beta})\right], 
\label{eqn:linresp}
\end{equation}
where $V_{\alpha}$ and $\theta_{\alpha}$ are the voltage and
temperature, respectively, at terminal $\alpha$, and $G_{\alpha\beta}$
and $L_{\alpha\beta}$ are the electrical conductance and
thermoelectric coefficients, respectively, between terminals $\alpha$
and $\beta$. The $L_{\alpha\beta}$ are directly related to the
thermopower, or Seebeck coefficients $\mathcal{S}_{\alpha\beta}\equiv
(V_{\alpha}-V_{\beta})/(\theta_{\alpha}-\theta_{\beta})|_{I=0}$,
through $L_{\alpha\beta} = -\sum_\gamma G_{\alpha\gamma}
\mathcal{S}_{\gamma\beta}$, where the sum runs over all terminals
\cite{Butcher1990}.

The transport properties in open mesoscopic systems can conveniently
be described by the scattering approach. The conductance coefficients,
see Eq.~\eqref{eqn:linresp}, are given by \cite{Buttiker1986}
\begin{equation}
G_{\alpha\beta}(B)=\frac{2e^2}{h}\int dE\left(-\frac{df(E)}{dE}\right)T_{\alpha\beta}(E,B)\,,
\end{equation}
for $\alpha \neq \beta$, where $T_{\alpha\beta}(E,B)$ is the
transmission function describing scattering of particles at energy $E$
from terminal $\beta$ to $\alpha$, $B$ is the magnetic field, and
$f(E)$ is the equilibrium Fermi distribution
function. Correspondingly, the thermoelectric coefficients are given
by \cite{Butcher1990}
\begin{equation}\label{eqn:Lalphabeta}
L_{\alpha\beta}(B)=\frac{2e}{h\theta_0}\int dE E\left(-\frac{df(E)}{dE}\right)T_{\alpha\beta}(E,B),
\end{equation}
where we have set the background chemical potential to zero.

Microreversibility demands that the transmission function obeys the
magnetic-field symmetry
$T_{\alpha\beta}(E,B)=T_{\beta\alpha}(E,-B)$. The resulting symmetry
for the conductance, $G_{\alpha\beta}(B)=G_{\beta\alpha}(-B)$, has
been thoroughly investigated over the last few decades
\cite{Buttiker1986,Benoit1986, vanHouten1989}. Since the
thermoelectric coefficients depend directly on the transmission
function, $L_{\alpha\beta}$ should obey the same symmetry properties
as $G_{\alpha\beta}$. This gives, writing out the diagonal and
off-diagonal relations separately,
\begin{equation}
L_{\alpha\alpha}(B)=L_{\alpha\alpha}(-B), \hspace{0.5cm} L_{\alpha\beta}(B)=L_{\beta\alpha}(-B)\,.
\label{eqn:Lmatsym}
\end{equation}

To experimentally test these symmetries, we first determine
$G_{\alpha\beta}$ through electric bias measurements with no thermal
bias ($\Delta\theta_{\alpha\beta} = 0$) (see Appendix B). Thereafter,
the thermoelectric coefficients are investigated by thermally biasing
the system under zero-electric-current conditions (with floating
terminals), measuring the resulting potentials in all reservoirs, and
using Eq.~\eqref{eqn:linresp} as explained in the following.

The induced temperature increase at terminal $\alpha$ can be written
as a Fourier sum, $\Delta \theta_{\alpha}(t)\equiv
\theta_{\alpha}(t)-\theta_0=\sum_{n=0}\Delta
\theta_{\alpha}^{(n)}\sin(n\omega t)$, where $\omega$ is the frequency
of the heating current. This allows us to write the different Fourier
components of the linear response current expression in
Eq.~\eqref{eqn:linresp} as
\begin{equation}
0=\sum_{\beta}\left[G_{\alpha\beta}V_{\alpha\beta}^{(n)}+L_{\alpha\beta}\theta_{\alpha\beta}^{(n)}\right], 
\label{eqn:linrespTB}
\end{equation}
where $V_{\alpha\beta}^{(n)}\equiv V_{\alpha}^{(n)}-V_{\beta}^{(n)}$,
and $\theta_{\alpha\beta}^{(n)}\equiv
\theta_{\alpha}^{(n)}-\theta_{\beta}^{(n)}
=\Delta\theta_{\alpha}^{(n)}-\Delta\theta_{\beta}^{(n)}$. When using
Joule heating (quadratic in heating current), one expects the second
harmonic to give the strongest contribution to the thermoelectric
response. Indeed, in our experiment we find that $n = 2$ gives the
largest signal and provides the clearest data to determine the range
of the linear response regime; in the following, we only consider the
second harmonics in Eq.~\eqref{eqn:linrespTB}.  Heating the
$\gamma$'th terminal, and making the assumption that the unheated
terminals remain cold, we can make use of the relation
$\sum_{\alpha}L_{\alpha\beta}=\sum_{\beta} L_{\alpha\beta}=0$, which follows from
the unitarity of the scattering matrix \cite{Butcher1990}, and write
\begin{equation}
L_{\alpha\gamma}\Delta\theta_{\gamma}^{(2)}=\sum_{\beta} G_{\alpha\beta}V_{\alpha\beta}^{(2)}\,.
\label{eqn:Lmateq}
\end{equation}
Here, $V_{\alpha\beta}^{(2)}$ represents the measured values when
heating terminal $\gamma$. Since the $G_{\alpha\beta}$ elements depend
weakly on magnetic field up to $B\sim 50$ mT  (see Appendix B), we can use
Eq.~\eqref{eqn:Lmateq} and $G_{\alpha\beta}$ to test the
magnetic-field symmetries of
$L_{\alpha\gamma}\Delta\theta^{(2)}_\gamma$ directly by analysing the
$B$-field dependence of the measured $V_{\alpha\beta}^{(2)}$. In the
following, we also assume that $\Delta\theta^{(2)}_\gamma$ is
independent of $B$, so that all of the $B$-field dependence in
$L_{\alpha\gamma}\Delta\theta^{(2)}_\gamma$ comes from
$L_{\alpha\gamma}$.
 
\begin{figure}
\centerline{\psfig{figure=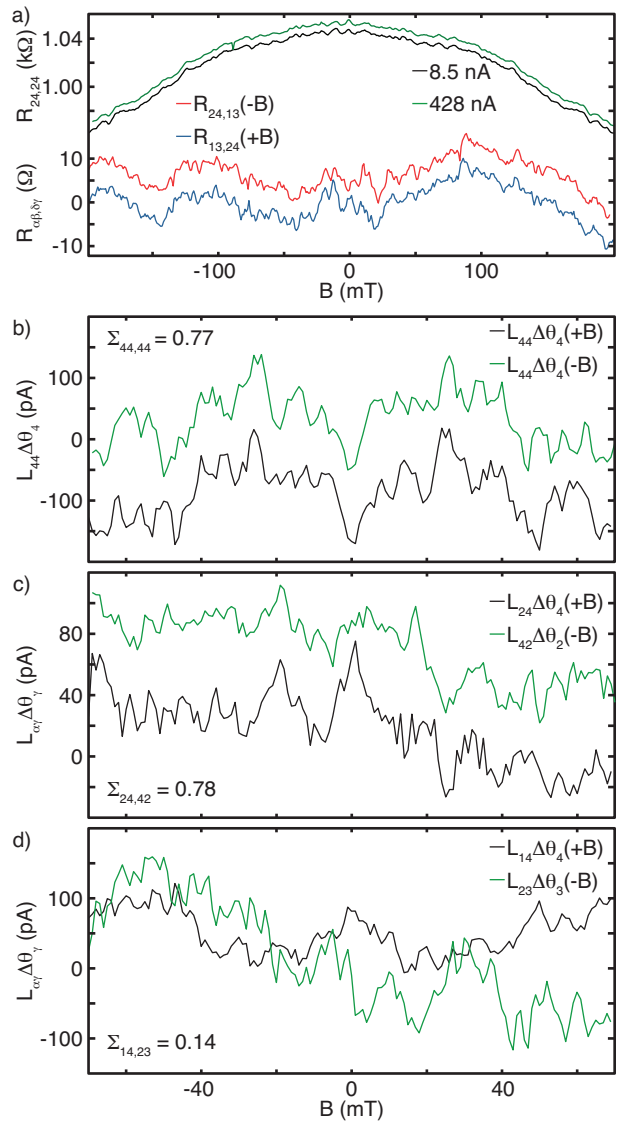}}
\caption{(Color online) a) Representative four-terminal resistances
  $R_{\alpha\beta,\gamma\delta}(B)$ as a function of magnetic
  field. The symmetry
  $R_{\alpha\beta,\gamma\delta}(B)=R_{\gamma\delta,\alpha\beta}(-B)$
  is clearly visible; small deviations are discussed in Appendix
  B. Note that the resistance $R_{24,24}(B)$ depends weakly on
  magnetic field for $-50~\mbox{mT}<B<50~\mbox{mT}$. b)-d) Magnetic
  field traces of the thermoelectric coefficient
  $L_{\alpha\beta}(B)$. Each panel also displays the corresponding
  quantitative symmetry parameter $\Sigma_{\alpha\beta,\gamma\delta}$
  [Eq. (\ref{eqn:S})] calculated for the range
  $-50~\mbox{mT}<B<50~\mbox{mT}$, for the respective pair of traces
  shown. The symmetry of the diagonal terms
  $L_{\alpha\alpha}(B)=L_{\alpha\alpha}(-B)$ is clearly visible in b),
  as well as that of the off-diagonal terms
  $L_{\alpha\beta}(B)=L_{\beta\alpha}(-B)$, in c). For comparison, an
  example of the expected absence of symmetries, here between
  $L_{14}(B)­$ and $L_{23}(-B)$, is illustrated in d) and manifested
  by a $\Sigma_{14,13}$ value near zero. The green and red curves in
  a), b), and c) are offset for clarity. All measurements were
  performed at a cryostat temperature of $\theta_0 = 240$ mK.}
\label{fig:fig2}
\end{figure}

\section{Thermoelectric reciprocity relations}

The symmetry relations predicted by Eq.~\eqref{eqn:Lmatsym} are 
clearly visible in the representative magnetic field traces for 
$L_{\alpha\gamma}\Delta\theta^{(2)}_\gamma$ presented in 
Fig.~\ref{fig:fig2}b),c). We also find that there is no significant symmetry 
relation between $L_{\alpha\beta}(B)$ and $L_{\gamma\delta}(-B)$ for 
$\alpha\beta \neq \gamma\delta$  (see Fig.~\ref{fig:fig2}d) for an example). 
The $L_{\alpha\gamma}\Delta\theta^{(2)}_\gamma$ typically oscillate 
around zero, a signature of quantum interference effects \cite{Esposito1987, 
Molenkamp1992,Langen1998}. 

To quantify the degree of symmetry, we make use of the correlation
coefficient \cite{pearson} between $L_{\alpha\beta}(B)$ and
$L_{\gamma\delta}(-B)$ (see Appendix C). We first introduce the normalized
thermoelectric coefficients
\begin{equation}
\mathcal{L}_{\alpha\beta}(B) \equiv \frac{L_{\alpha\beta}(B) - \langle L_{\alpha\beta}(B)\rangle}{\sqrt{\langle [L_{\alpha\beta}(B)]^2\rangle - \langle L_{\alpha\beta}(B)\rangle ^2}}\,,
\label{eqn:normLab}
\end{equation}
where $\langle ... \rangle $ denotes the average over magnetic fields
from $-50$ mT to $50$ mT. This magnetic field range was chosen to
avoid the onset of classical commensurability effects. We calculate
Eq.~\eqref{eqn:normLab} using $L_{\alpha\beta}\Delta\theta^{(2)}$ in
place of $L_{\alpha\beta}$, since we have assumed that
$\Delta\theta^{(2)}$ is independent of $B$ and thus cancels out. We
then define the symmetry parameter as 
\begin{equation} \label{eqn:S}
\Sigma_{\alpha\beta,\gamma\delta} \equiv \langle \mathcal{L}_{\alpha\beta}(B)\mathcal{L}_{\gamma\delta}(-B)\rangle \,.
\end{equation}
Note that $\Sigma_{\alpha\beta,\gamma\delta}$ goes from $+1$ for
complete symmetry, to $-1$ for complete anti-symmetry. We stress that
$\Sigma_{\alpha\beta,\gamma\delta}$ is well suited to quantify the
overall symmetry of functions which, like the ones in
Fig. \ref{fig:fig2}, display rapid oscillations on top of a smooth,
slowly oscillating background. The need for a quantitative symmetry
measure becomes apparent when comparing Figs. \ref{fig:fig2}b) and
\ref{fig:fig2}c); while the corresponding symmetry parameters are essentially identical,
$\Sigma_{44,44}\approx \Sigma_{24,42}$, to the bare eye the curves in
Fig. \ref{fig:fig2}b) appear more symmetric than the ones in
\ref{fig:fig2}c). We have analyzed all combination of curves
$\alpha\beta,\gamma\delta$ and also compared the results to another
potential symmetry measure, the magnitude of the fluctuation of the
difference, $ \langle[
\mathcal{L}_{\alpha\beta}(B)-\mathcal{L}_{\gamma\delta}(-B)]^2\rangle$. The
result (not presented here) firmly establishes that
$\Sigma_{\alpha\beta,\gamma\delta}$ is a reliable symmetry measure.

Deviations from the perfect symmetries predicted in
Eq.~\eqref{eqn:Lmatsym} are seen in our measurements. To rule out
noise as the cause of the limited symmetries, we verified that two
traces measured almost two weeks apart showed very high correlation
(see Appendix C), demonstrating the high repeatability of these
fluctuations. We attribute the limited symmetry in Fig. 2b),c) mainly
to the same mechanisms that limit the observed conductance
symmetry. In addition, however, we offer two other possible
mechanisms: i) the unheated terminals do not remain cold, which would
modify Eq.~\eqref{eqn:Lmateq}; and ii) inelastic scattering, which at
finite temperature can lead to asymmetries in the thermopower even in
the linear response regime \cite{Sanchez2011, Saito2011,Entin2012}.

\section{Symmetry breakdown}

\begin{figure}
\centerline{\psfig{figure=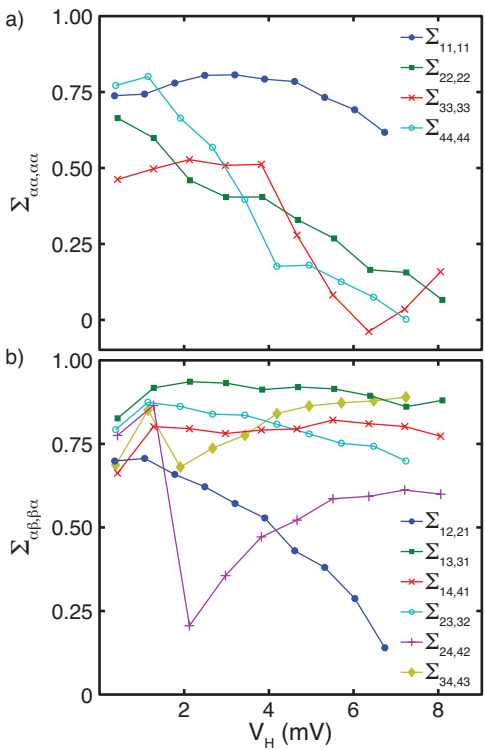}}
\caption{(color online) Heating voltage dependence of the symmetries described in Eq.~\eqref{eqn:Lmatsym} for the a) diagonal, $\alpha = \beta$, and b) off-diagonal, $\alpha \neq \beta$, elements of $L_{\alpha\beta}$. In panel a), a clear trend for decreasing symmetry with increasing thermal bias is seen for the diagonal elements. This same trend is only present in three of the six curves in panel b).}
\label{fig:fig3}
\end{figure}

One can expect the symmetry of $L_{\alpha\beta}$ to break down for
finite heating voltage, analogous to the well-established breakdown of
symmetries in the differential conductance \cite{Sanchez2004,
  Spivak2004} observed at finite bias voltage in mesoscopic systems
\cite{Linke1998, Marlow2006, Leturcq2006, Zumbuhl2006, Angers2007,
  Hartmann2008}. In Fig.~\ref{fig:fig3}, all ten symmetry relations
defined by Eq.~\eqref{eqn:Lmatsym} are plotted as a function of
heating voltage $V_H$. At low $V_H$, all symmetries described by
Eq.~\eqref{eqn:Lmatsym} manifest themselves, with
$\Sigma_{\alpha\beta,\beta\alpha}\gtrsim0.5$. As $V_H$ is increased
though, the trend in the diagonal elements, $\alpha = \beta$, is
towards decreased symmetry, while the off-diagonal elements, $\alpha
\neq \beta$, remain fairly symmetric with a slight trend to decrease.

The overall tendency is for the $B$-field symmetries of
$L_{\alpha\beta}$ to be suppressed with increasing thermal bias. From
further analysis of our measured data (see Appendix D), we establish that the
linear-response regime extends to about $V_H \approx 1$ mV.  The
decreasing symmetry observed in $L_{\alpha\alpha}$,
Fig.~\ref{fig:fig3}a), is then consistent with symmetry-breaking due
to non-linear thermoelectric behavior \cite{san13,mea13,sun13},
analogous to non-linear electronic effects. Increased inelastic
scattering due to heating effects may also play a role. 

Further theoretical as well as experimental investigations are needed
to explain the observed difference in symmetry between diagonal and
off-diagonal elements of $L_{\alpha\beta}$ for $V_H \geq 1$mV, see
Fig. \ref{fig:fig3}. In particular, in contrast to linear response
theory \cite{Butcher1990}, existing non-linear theory
\cite{san13,mea13,sun13} does not predict any simple relations between
diagonal and off-diagonal elements.  

\section{Conclusion}
In conclusion, we have verified that the TE reciprocity relations
predicted more than 20 years ago \cite{Butcher1990} manifest
themselves in a mesoscopic device in the linear transport regime. The
relations were observed at low temperatures, where inelastic
scattering (predicted to suppress the symmetries
\cite{Sanchez2011,Saito2011,Entin2012}) can be expected to be small.
At finite thermal bias we observe a breakdown of the reciprocity
relations, tentatively due to a combination of inelastic scattering
and non-linear thermal transport. Further investigations are needed to
quantify the robustness of the reciprocity relations with respect to
these mechanisms. Of particular interest will be the role of sharp
features in the transmission function or density of states commonly
used for energy filtering to enhance thermoelectric performance, for
example in low dimensional coolers \cite{Giazotto2006, Edwards1993}
and highly efficient thermoelectric generators \cite{Humphrey2002,
  Hicks}. The possibility to experimentally control the absence or
presence of the TE symmetry relations opens for exciting and
fundamentally new opportunities in increasing TE energy efficiency
\cite{Casati2011,Brandner,Bala}.

\section{Acknowledgements} 
We acknowledge financial support from NSF IGERT grant No.~DGE-0549503,
the National Science Foundation Grant No.~DGE-0742540, ARO Grant
No. W911NF0720083, nmC@LU, the ESF Research Network EPSD, the
Foundation for Strategic Research (SSF), the Swedish Energy Agency
(project number 38331-1), and MINECO Grant No. FIS2011-23526. Effort
sponsored by the Air Force Office of Scientific Research, Air Force
Material Command, USAF, under grant number FA8655-11-1-3037. The
U.S. Government is authorized to reproduce and distribute reprints for
Governmental purposes notwithstanding any copyright notation thereon.

\appendix
\section{Measurement of transport coefficients}

We individually use electric or thermal biases to determine the
elements of the electrical conductance matrix, $G_{\alpha\beta}$, and
the thermoelectric matrix, $L_{\alpha\beta}$, respectively. Electric
biases are generated by applying a 37 Hz drive current between any two
of the four central terminals. The measured $G_{\alpha\beta}$ are
discussed below. A representative circuit configuration used for
thermal-bias measurements is shown to the left in Fig.~1 in the main
text. The thermal bias is generated by individually heating one
terminal by applying two $180^\circ$ out-of-phase, 37 Hz voltages,
denoted by $\nu_H^\pm$, to the two channel contacts (in Fig.~1, main
text, contacts A and B are used to heat terminal 1). This heating
configuration is designed to eliminate any electric bias of the
terminal due to the heating current. Additionally, a DC shift was
applied to the thermal bias to cancel out residual DC offsets measured
at the respective terminal's voltage probe. We note that only one
terminal is heated at a time, and that negligible electric current is
drawn through the junction during thermovoltage measurements. For
further details on the thermal bias measurements, see
Ref.~[\onlinecite{Matthews2012}]. Under each type of bias, all four-terminal
voltages, $\{V_1, V_2, V_3, V_4\}$, were simultaneously measured using
lock-in detection. 

For thermal biasing, one pad was electrically heated using AC heating
currents of between $300-400$ nA at a frequency of $37$ Hz. The
externally measured, two-terminal resistance of heating pads was
between $1000-1400$ $\Omega$, such that $1$k$\Omega$ is a reasonable
upper limit of the resistance of the heating pad itself. This gives an
upper limit of the electric heating power delivered to a heating pad
of about $0.1$ nW. Finite-element simulations (COMSOL) that use the
actual device geometry and take into account heat leaks to the
surrounding 2DEG (but neglect additional heat leaks to the phonons)
predict an upper limit of a temperature rise $\theta \approx 0.5
-1$K. Using the upper limit $\theta=1$ K, and estimating the thermal
conductance between two heating pads (based on the electrical
resistance between two pads of typically $1$ k$\Omega$ and using the
Wiedemann-Franz law), and again neglecting other heat leaks such as
electron-phonon coupling, an upper limit for the resulting heating
power to neighboring heat pads is about $6 \times 10^{-3}$ nW, about
100 times smaller than the intentional heating.

\section{Conductance measurements}

\begin{figure}
\centerline{\psfig{figure=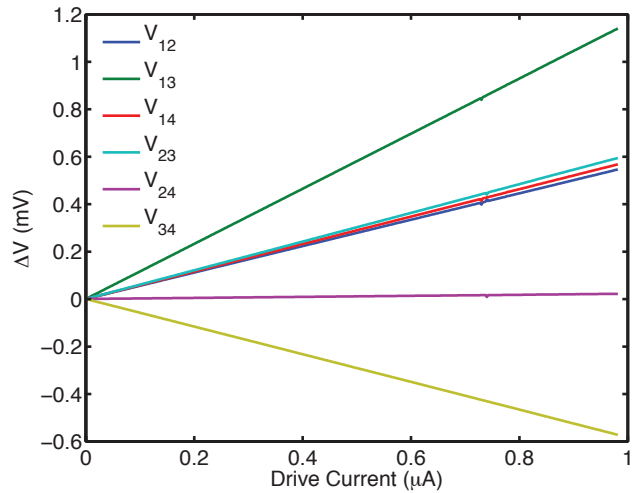}}
\caption{The induced voltages $V_{\alpha\beta}$ between all pair of
  terminals $\alpha$ and $\beta$, as a function of an electrical
  current bias between terminal 1 and 3.}
\label{SIfig1}
\end{figure}

To determine the conductance coefficients at $B=0$, three different
biasing configurations were utilized; each had the 37 Hz bias current
injected at terminal 1, which was then extracted at terminals 2, 3 and
4 for the three biasing configurations. The rms amplitude of the bias
current, $I$, was varied from 4 to 980 nA to check for linearity in
the current-voltage characteristics, sec Fig. \ref{SIfig1} for a set
of representative curves.

%
We stress that for the entire bias current range used here, the
deviations from a linear-in-current voltage response are negligibly
small, three orders of magnitude smaller than the linear
response. 

In the linear response regime, only the first harmonic of the voltage
response is significant, and we can write the current as
\begin{equation}
I_{\alpha}^{(1)}=\sum_{\beta\neq \alpha}G_{\alpha\beta}V_{\alpha\beta}^{(1)}\,, \hspace{0.5cm} V_{\alpha\beta}^{(1)}=V_{\alpha}^{(1)}-V_{\beta}^{(1)}\,,
\label{eqn:linrespCB}
\end{equation}
where $I^{(1)}_{\alpha}=+(-)I$ at the injection (extraction) terminal
and zero at the two floating terminals; and the superscript denotes
the harmonic of the drive voltage frequency. Due to current
conservation and gauge invariance, the conductances obey the sum rules
$\sum_{\alpha}G_{\alpha\beta}=\sum_{\beta}G_{\alpha\beta}=0$. Measurements
of the $1\omega$ voltage responses, $V_\alpha^{(1)}$, at $B=0$ for
each of the three biasing configurations, together with the sum rules,
allow us to determine all 16 elements $G_{\alpha\beta}$ of the conductance matrix $G$. Our
analysis gives
\begin{equation}
G(B=0)=\frac{2e^2}{h}\left(\begin{array}{cccc}
-21.3 & 11.7 & 1.5 & 8.1 \\
11.5 & -25.3 & 10.5 & 3.3 \\
1.6 & 10.4 & -20.1 & 8.1 \\
8.1 & 3.2 & 8.1 & -19.5 \end{array} \right)\,.
\label{eqn:condmat}
\end{equation} 
All conductance coefficients obey $G_{\alpha\beta} >2e^2/h$,
demonstrating that electron transport is in the open regime.

Eq. ~\eqref{eqn:condmat} clearly shows that $G_{\alpha\beta}
\approx G_{\beta\alpha}$, as expected from the fundamental symmetry of
$G$. However, microreversibility predicts a complete symmetry. To
further investigate the small asymmetry apparent in
Eq.~\eqref{eqn:condmat} we plot in Fig.~2a) in the main text the
four-terminal resistances
$R_{\alpha\beta,\gamma\delta}(B)=V_{\alpha\beta}^{(1)}/I_{\gamma\delta}$,
with $\gamma(\delta)$ representing the current injection (extraction)
terminal, as a function of magnetic field, $B$. The fluctuations
observed in these traces are due to wave interference effects typical
for open mesoscopic conductors. We see that the required symmetry
relation
$R_{\alpha\beta,\gamma\delta}(B)=R_{\gamma\delta,\alpha\beta}(-B)$ is
slightly violated, in particular at the lower fields, $|B| \lesssim
50$ mT, consistent with the slightly asymmetric conductance
coefficients $G_{\alpha\beta}$ observed in Eq~\eqref{eqn:condmat}. The
origin of this asymmetry is not clear. We can rule out noise and
nonlinear effects \cite{Lofgren2006} as the cause by comparing to a
second measurement taken at much higher bias current, which essentially shows the same asymmetry. A
magnetic sample holder can also be ruled out, as care was taken to use
non-magnetic materials. Leakage currents due to the voltage probes are
also found to be negligibly small.  We speculate that magnetic
impurities may play a role.

\begin{figure}
\includegraphics[width=0.85\linewidth]{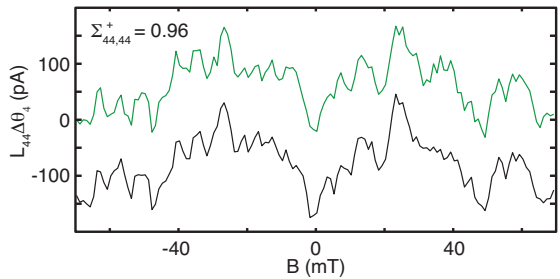}
\caption{Magnetic field traces of the thermoelectric coefficient $L_{44}(B)$ taken almost two weeks apart. The modified symmetry parameter $\Sigma_{44,44}^+$ (see text), calculated for the range $-50~\mbox{mT}<B<50~\mbox{mT}$, gives a value close to unity, demonstrating a high degree of repeatability. The measurements were performed at a cryostat temperature of $\theta_0 = 240$ mK.}
\label{fig:figSI}
\end{figure}
%

The typical magnitude of the magnetic field dependent
oscillations of $G_{\alpha\beta}$ can be qualitatively estimated from
corresponding fluctuations of the longitudinal four-terminal
resistance $R_{24,24}$ in Fig. 2 in the manuscript (similar results
obtained for other $R_{\alpha\beta,\alpha\beta}$, not presented). In
the magnetic field range $-50~\mbox{mT}<B<50~\mbox{mT}$ the
fluctuations are of the order of a few percent, to be expected
\cite{been} from a mesoscopic system in the open transport regime with
a typical conductance $G_{\alpha\beta}\sim 10e^2/h$, see
Eq. (\ref{eqn:condmat}).

\section{Symmetry measure and reproducibility}

To quantify the degree to which the two data sets $L_{\alpha\beta}$
and $L_{\gamma\delta}$ obey the symmetry relation
$L_{\alpha\beta}(B)=L_{\gamma\delta}(-B)$, we make use of Pearsons
product-moment correlation coefficient, or r-correlation coefficient
\cite{pearson}, between $L_{\alpha\beta}(B)$ and
$L_{\gamma\delta}(-B)$, our Eqs. (7) and (8) in the main text. The
r-correlation coefficient is a well established measure to quantify
the correlation between two data sets. Importantly, the r-correlation
coefficient is known to be a reliable measure of correlation in the
absence of outliers, i.e. extreme, isolated measurement points. We
carefully investigated our data to rule out such extreme points.

To investigate the repeatability of the magnetic field traces
$L_{\alpha\beta}(B)$, two traces measured almost two weeks apart are
shown in Fig. \ref{fig:figSI}. The degree of correlation between the
two traces is quantified by the modified symmetry parameter
$\Sigma_{\alpha\beta,\gamma\delta}^+=\sum_B\mathcal{L}_{\alpha\beta}(B)\mathcal{L}_{\gamma\delta}(B)$,
reaching $1$ for perfect correlation
$\mathcal{L}_{\alpha\beta}(B)=\mathcal{L}_{\gamma\delta}(B)$.

\section{Linear thermal bias response}
To establish the range over $V_H$ where we expect 
a linear-in-temperature response, we used the solution of the 
quasi-one-dimensional heat diffusion equation \cite{Giazotto2006} to 
estimate the temperature rise in the heated channel, labelled below as 
the $\alpha$'th terminal, as a function of $V_H$,
\begin{equation}\label{eqn:Temperature}
\theta_\alpha = \theta_0\sqrt{1+\left(\frac{V_H}{V_C}\right)^2\cos^2(\omega t)}\,,
\end{equation}
where $V_C$ is a heating channel dependent parameter. Using
Eq.~\eqref{eqn:Temperature}, we can estimate the predicted Fourier
components of $L_{\alpha\gamma}\Delta\theta^{(2)}_{\gamma}$ and
compare them to our measured data,
$\sum_{\beta}G_{\alpha\beta}V_{\alpha\beta}^{(2)}$. In this way, we
have clearly established that the linear-response regime extends to
about $V_H \approx 1$ mV, which corresponds to
$\Delta\theta^{(2)}_\gamma/\theta_0\approx 0.21$ to $0.75$ depending
on which terminal is heated.

\bibliography{RefsPRL}

\end{document}